\documentclass[11pt, a4paper]{article}
\usepackage{jheppub}
\usepackage{bbm}
\usepackage{comment}
\usepackage{amsmath}
\usepackage{amssymb}
\usepackage{amsfonts}
\usepackage{mathrsfs}
\usepackage{graphicx}
\usepackage[11pt]{moresize}
\usepackage[normalem]{ulem}
\usepackage[dvipsnames]{xcolor}
\usepackage[]{hyperref}
\usepackage{orcidlink}
\usepackage{relsize}
\usepackage{adjustbox}
\usepackage[verbose]{placeins}
\usepackage{multirow}
\usepackage{float}
\usepackage{footnote}
\usepackage{tablefootnote}
\usepackage{lineno}
\usepackage{makecell}
\usepackage{subcaption}

\hypersetup
{
 bookmarks  = true,           
 unicode    = false,          
 pdfcreator = {RevTeX},       
 colorlinks = true,           
 linkcolor  = blue,            
 citecolor  = blue,           
 filecolor  = black,          
 urlcolor   = blue,           
}

\renewcommand{\mp}{m_\phi}
\newcommand{\E}{E_\nu}

\title{Probing Scalar-Induced First-Order Phase Transitions with the Diffuse Supernova Neutrino Background}

\author[a] {\orcidlink{0000-0002-5508-7751} Sudipta Das,}
\author[b] {\orcidlink{0009-0004-6370-5419} Shamik Niyogi,}
\author[b] {\orcidlink{0000-0001-7948-4332} Manibrata Sen}

\affiliation[a]{ Department of Physics and Astronomy, University of Iowa, Iowa City, IA 52242, USA}
\affiliation[b]{Department of Physics, Indian Institute of Technology Bombay, Powai, Mumbai 400076, India}

\emailAdd{sudipta-das@uiowa.edu}
\emailAdd{shamik$\_$niyogi@iitb.ac.in}
\emailAdd{manibrata@iitb.ac.in}

\abstract{We investigate the possibility of probing a scalar-induced first-order phase transition (FOPT) inside core-collapse supernovae through the diffuse supernova neutrino background (DSNB). We consider a light scalar coupled to neutrinos that is produced and thermalised in the proto-neutron-star core. This leads to two distinct neutrino components: a prompt contribution from thermally produced scalars and a delayed, spectrally narrow contribution from the decay of coherent scalar excitations due to the FOPT. We study the resulting DSNB signal at Super-Kamiokande, Hyper-Kamiokande, and DUNE, finding a strong dependence on the neutrino mass ordering. The recent Super-Kamiokande data already constrain the scalar parameter space for inverted ordering and quasi-degenerate neutrino masses, while future detectors can significantly improve this sensitivity. Our results demonstrate that DSNB measurements can provide a novel probe of finite-temperature scalar dynamics in core-collapse supernovae.}

\begin{document}
\maketitle
\section{Introduction}
\label{sec:intro}
Core-collapse supernovae (CCSNe) are among the most energetic events in the Universe and represent one of the few astrophysical environments where matter reaches extreme densities and temperatures. During the gravitational collapse of the iron core of a massive star, approximately $10^{53}$ ergs of gravitational binding energy is released, of which nearly $99\%$ is carried away by neutrinos over a timescale of around ten seconds. Since neutrinos escape from regions that are opaque to electromagnetic radiation, they provide a unique window into the physical conditions deep inside the proto-neutron star (PNS). Consequently, supernova (SN) neutrinos have long served as powerful probes of neutrino properties, dense matter and the presence of feebly interacting particles in extensions of the Standard Model (SM).

The only direct observation of SN neutrinos to date came from SN1987A, whose detection established the basic theoretical picture of neutrino emission from collapsing stars~\cite{Kamiokande-II:1987idp,PhysRevLett.58.1494,ALEXEYEV1988209,Dodelson:1992tv,annurev:/content/journals/10.1146/annurev.ns.40.120190.001145}. However, galactic (or near-galactic) CCSNe are extremely rare, which prevents us from exploring this laboratory further. A complementary observable is provided by the Diffuse Supernova Neutrino Background (DSNB)~\cite{Lunardini:2009ya, Lunardini:2010ab,Beacom:2010kk,Lunardini:2012ne}, which consists of the integrated neutrino emission from all past core-collapse events throughout cosmic history. Since every successful and failed core-collapse contributes to the DSNB, its flux carries cumulative information about the cosmic SN rate, stellar evolution, black-hole formation, and the neutrino emission mechanisms operating inside individual explosions. Any new physics that systematically modifies neutrino production in CCSNe is therefore expected to leave an observable imprint on the DSNB~\cite{Chakraborty:2008zp,Farzan:2014gza,Moller:2018kpn, DeGouvea:2020ang,Das:2021lcr,Das:2022xsz, deGouvea:2022dtw, Ivanez-Ballesteros:2022szu, Akita:2022etk,Suliga:2021hek,JUNO:2022lpc,Balantekin:2023jlg,MacDonald:2024vtw, Das:2024ghw, Roux:2024zsv,Perez-Gonzalez:2025qjh,Granelli:2026bem,Dighe:2026mxk}.

The experimental search for the DSNB has advanced rapidly in recent years~\cite{Horiuchi:2008jz,Lunardini:2010ab,Sawatzki:2020mpb}. The incorporation of gadolinium into the Super-Kamiokande~(Super-K) detector has significantly improved neutron tagging and background rejection, leading to the most stringent searches for the DSNB to date~\cite{Beacom:2003nk,Super-Kamiokande:2008mmn,Laha:2013hva}. Very recently, the Super-K collaboration reported its results 
from 956.2 days of data, showing an $\sim$ 1.2$\sigma$ disagreement with the null DSNB hypothesis~\cite{Super-Kamiokande:2025sxh}. Future experiments, including Hyper-Kamiokande~(Hyper-K)~\cite{Hyper-Kamiokande:2018ofw}, Deep Underground Neutrino Experiment~(DUNE)~\cite{DUNE:2020lwj} and Jiangmen Underground Neutrino Observatory~(JUNO)~\cite{JUNO:2015zny}, are expected to improve the sensitivity considerably through larger detector volumes and complementary detection~\cite{Moller:2018kpn}. 

Among the many proposed extensions of the SM, light scalar fields constitute a particularly well-motivated possibility. Such scalars arise naturally in numerous theories of neutrino mass generation~\cite{Cai:2017jrq,deGouvea:2016qpx,Mandal:2022zmy,Sen:2023uga, He:1998ng}, hidden dark sectors~\cite{Bertacca:2010ct,Arbey:2006it,Mishra:2018tki,Gao:2009me,Arun:2017uaw}, neutrinophilic interactions~\cite{Babu:2019iml,Brdar:2020nbj,DeGouvea:2019wpf,Arguelles2023,Berryman:2022hds,deLima:2026fsy,Dev:2024ygx,Das:2025zts,Dutta:2025rxh} and scalar portal models~\cite{Arcadi:2019lka,Arcadi:2021mag,Boiarska:2019jym,Lebedev:2021xey}, to name a few. If their masses lie in the tens of MeV ballpark range, and their couplings to neutrinos are sufficiently large, they can be abundantly produced inside the hot and dense environment of a PNS through their interactions with neutrinos. This can be expected to modify the dynamics of a CCSN. 

The presence of these scalars can be constrained if they free-stream out of the PNS after production, thereby leading to additional cooling of the SN core and modifying the energy transport inside the SN~\cite{Raffelt:1996wa, Raffelt:2006cw, DiLuzio:2021ysg, Heurtier:2016otg, Brune:2018sab, Caputo:2021rux, Caputo:2022rca, Fiorillo:2025yzf}. Furthermore, if these scalar particles additionally decay into neutrinos, they can produce characteristic distortions in the neutrino spectrum~\cite{ Fiorillo:2022cdq, Akita:2022etk, Akita:2022etk, Telalovic:2024cot}. However, if the scalars are trapped inside the SN core, cooling is inefficient~\cite{Fiorillo:2023cas}. Rather than acting as a freely escaping particle, the scalar thermalises with the neutrino bath through repeated scattering and annihilation processes and becomes an integral part of the thermal plasma~\cite{Davoudiasl:2005fd, Davoudiasl:2025cbo}. Consequently, its evolution is governed not only by its vacuum properties but also by the thermodynamic evolution of the SN core. 

In such a situation, finite-temperature effects can substantially modify the scalar potential inside the PNS. At sufficiently high temperatures, thermal corrections can restore the symmetry of a scalar potential that is spontaneously broken in vacuum. As neutrino emission cools the PNS, the temperature eventually falls below the critical temperature of the effective potential. The scalar field then undergoes a transition back to the broken phase. If the finite-temperature effective potential contains a barrier separating the symmetric and broken minima, this transition proceeds through a first-order phase transition (FOPT). During this transition, the vacuum energy released is converted into coherent oscillations of the scalar field about its true vacuum.

In this scenario, the scalar field can manifest itself in two distinct ways. Scalar particles are produced inside the SN core due to usual quantum excitations of the scalar field about its vacuum. However, there exists another possibility. Coherent oscillations of the scalar field about the true vacuum during the SN cooling can also create a scalar condensate, similar to a reheating scenario. The crucial part is that both the realisations of the scalar field will decay to neutrinos, with very different signatures. 

The first originates from thermal scalar particles that decouple from the PNS after their optical depth falls below unity. These free-streaming scalars subsequently decay into neutrinos outside the star, generating a \emph{prompt} contribution to the neutrino spectrum. Since the scalars obey Bose-Einstein statistics, the resulting prompt neutrino spectrum possesses a characteristic shape distinct from the conventional SN emission. 

The second contribution arises from the coherent scalar oscillations around the true vacuum. The energy released in the transition is transferred into neutrinos through the scalar decay. Assuming the scalar is non-relativistic, the decay produces neutrinos with energies around half the scalar mass. This leads to a \emph{delayed} spectral component that is considerably narrower than the thermal SN neutrino spectrum.  Unlike the prompt signal, the delayed component directly probes the vacuum dynamics associated with the phase transition itself.

The existence of these two additional neutrino components modifies both the neutrino emission from an individual CCSN and, hence, the resulting DSNB. Since the DSNB integrates the contribution from every core-collapse event across cosmic history, even relatively small changes in the neutrino spectrum of an individual SN can give measurable effects in the integrated flux. The DSNB therefore provides a sensitive probe of scalar-induced phase transitions occurring inside PNS.

We investigate this possibility in this paper. We consider a generic light scalar coupled to neutrinos that remains thermalised inside the PNS and undergoes a finite-temperature FOPT during the cooling evolution of the SN. We describe the thermal production and transport of the scalar field and 
identify the conditions under which thermal effects can restore the
symmetry inside the core. Assuming a first-order transition, the prompt and delayed neutrino spectra arising from the thermal scalar population and the condensate are computed separately and their contribution to the DSNB are evaluated. We find that using the data reported by Super-K, we can already probe a significant portion of the parameter space of the underlying theory. Finally, we discuss the observational prospects of these signatures in upcoming neutrino experiments like Hyper-K and the DUNE.

The paper is organised as follows: section~\ref{sec:FOPT} describes the scalar dynamics inside the SN and the
phenomenological assumptions underlying the first-order phase
transition. Section~\ref{sec:Fluxes} discusses the neutrino flux arising from scalar decay to neutrinos within the SN. In section~\ref{sec:DSNB}, we briefly review the standard DSNB formulation, and show the modification in the DSNB due to the contributions from the scalar decay. Section~\ref{sec:Events_analysis} estimates the event rate at Super-K, Hyper-K, and DUNE. We derive the sensitivities of the three setups for the scalar induced FOPT in SN in section~\ref{sec: sensitivity}. Finally, we conclude in section~\ref{sec:conclusion}.

\section{Scalar induced phase transition inside supernova}
\label{sec:FOPT}
We consider a generic extension of the SM, consisting of a scalar, $\phi$, coupled to neutrinos. The relevant part of the Lagrangian is given by,
\begin{equation}
     \mathcal{L}\supset  y\,\bar{\nu}_L\nu_R \phi + {\rm h.c.} + m_R \overline{\nu^c_R} \nu_R - V(\phi) 
\end{equation}
where $V(\phi)$ is the potential associated with the scalar particle. Rather than focusing on a particular ultraviolet completion, our main objective is to study the phenomenological consequences of neutrinos coupled to a scalar whose finite-temperature dynamics become relevant inside a CCSN. In particular, we assume that
the scalar is sufficiently strongly coupled to the neutrino bath to be
produced and thermalised in the PNS, and that its finite-temperature
effective potential admits a first-order phase transition at a
temperature relevant to the PNS evolution.

The scalar potential at zero temperature can be parameterised as
\begin{equation}
    V(\phi) = -\mu^2\phi^2+\alpha\phi^3+\lambda\phi^4\,
\end{equation}
where $\mu$ is the coefficient of the quadratic term, $\alpha$ is a dimensional coupling characterising the cubic term which is crucial for a first order phase transition, and $\lambda$ is a dimensionless quartic coupling. For $\alpha=0$, the potential is invariant under a $Z_2$ transformation $\phi\rightarrow-\phi$, and the negative quadratic term leads to spontaneous symmetry breaking with a nonzero vacuum expectation value $\langle\phi\rangle=v_\phi$. A nonzero cubic term explicitly breaks this $Z_2$ symmetry and can contribute to the
appearance of a barrier between the symmetric and broken phases at
finite temperature.

This leads to the generation of active neutrino masses via the seesaw mechanism, where $m_\nu\approx (y\,v_\phi)^2/m_R$. 
This can be easily generalised to three active neutrinos and two right handed neutrinos, which is the most minimal extension required to generate observed active neutrino mass-squared differences.
As a representative benchmark, considering $m_R=100\,\mathrm{MeV}$ and $v_\phi=50\,\mathrm{MeV}$ gives
$y\simeq 6\times10^{-5}$ for $m_\nu\simeq 0.1\,\mathrm{eV}$.
However, since our analysis depends only on the existence of the scalar field and its interaction with neutrinos, these model-dependent details play no role in the subsequent phenomenology. The mass of $\phi$ can be obtained as
\begin{equation}
    m_\phi^2=\frac{d^2 V(\phi)}{d \phi^2}\Big|_{\phi=v_\phi}=-2\mu^2+6\alpha v_\phi+12\lambda v_\phi^2\,.
\end{equation}

Neutrinos remain trapped inside the core due to weak interactions with the surrounding matter. The neutrino number density is large enough such that any scalar possessing an appreciable coupling to neutrinos can also be efficiently produced through neutrino scattering and annihilation processes. The dominant reaction responsible for producing scalars is neutrino annihilation, $\nu \overset{(\,-\,)}{\nu} \leftrightarrow \phi \phi $. Since the active neutrinos inside a CCSN form a highly degenerate Fermi gas, the interaction rate is dominated by their chemical potential which can be quite large, $\mu_{\nu_L} \sim 150-200$ MeV. As a result, the per-particle production rate can be approximated, in the high $\mu_{\nu_L}$ limit, as $\Gamma \approx y^4 \mu_{\nu_L}/\pi^3$. For a benchmark choice of $y = 10^{-4}$ and $\mu_{\nu_L} = 150$ MeV, this rate turns out to be $\sim 10^5~\mathrm{s}^{-1}$. Since this interaction timescale is significantly shorter than the $\sim 10\,$s CCSN duration, we expect a large population of scalar particles inside the medium.

Once produced, the scalar population must also undergo sufficiently frequent
interactions with the neutrino bath to establish kinetic and chemical
equilibrium. This is established through the process,  $\phi\nu_L\rightarrow\nu_R$, followed by prompt decay, $\nu_R\rightarrow\phi\nu_L$. The absorption and subsequent decay provide efficient momentum and energy exchange between the scalar and neutrino sectors. Since the relevant interaction rates scale as $y^2$, the corresponding mean free path can be much smaller than the characteristic size of the neutrinosphere, preventing the scalar population from free-streaming out of the core. Since the decay can exist only if $m_{\phi} < m_R$, we restrict our analysis to $m_{\phi} < m_R \simeq 100\,{\rm MeV} \lesssim \mu_{\nu_L}$.  Once local thermal equilibrium is established, the scalar distribution approaches the Bose-Einstein distribution.

Throughout the trapped region, the scalar therefore shares the same local temperature as the neutrinos and contributes to the thermodynamic properties of the medium in precisely the same manner as any other relativistic bosonic degree of freedom. Therefore, the transport of scalar particles is governed by the same principles that govern neutrino transport inside a CCSN. As the neutrino density decreases, its chemical potential falls off, and hence the kinematic threshold of the absorption channel ($\phi\nu_L\rightarrow\nu_R$), given by $E_\phi E_{\nu_L}\sim E_\phi \mu_{\nu_L} \gtrsim m_R^2/4$, becomes difficult to achieve. As a result, the process become kinematically unavailable, allowing the scalar to escape. The scalar therefore transitions from an
optically thick regime in the inner core to a free-streaming regime
outside the transport sphere.

For the benchmark SN profiles and couplings considered here,
the scalar transport sphere lies close to the neutrinosphere, as found
in previous analyses of scalar-neutrino interactions~\cite{Davoudiasl:2005fd}. After decoupling, the scalar distribution freezes out while retaining the thermal spectrum inherited from the decoupling surface. Since collisions become negligible beyond this point, the subsequent evolution is governed entirely by free streaming and particle decay. In the phenomenological analysis below, we therefore assume that the
couplings are sufficiently large for the scalar to attain local
thermal equilibrium in the inner PNS while still allowing it to
decouple at a finite transport radius. The detailed solution of the
scalar Boltzmann transport equation in a time-dependent PNS background
is beyond the scope of this work.

\subsection{Finite-temperature effective potential and symmetry restoration}
Since the scalar remains in thermal equilibrium with the local surroundings, its effective potential depends explicitly on the temperature of the PNS. These finite temperature corrections modify the zero-temperature potential to
\begin{equation}
     V(\phi,T)=(c\, T^2-\mu^2)\phi^2+\tilde{\alpha}(T)\phi^3+\tilde{\lambda}(T)\phi^4,
\end{equation}
where $c\approx \lambda$ is is the dominant thermal correction to the quadratic term\footnote{Here, we neglect the contribution of the fermion loop to the thermal mass, as the Yukawa coupling ($y \sim 10^{-4}$) is smaller compared to the scalar self-coupling ($\lambda \sim \mathcal{O}(1)$).}, and $\tilde{\alpha}(T), \tilde{\lambda}(T)$ denotes the corrections to coefficients, which depend on the particle content and the couplings of the underlying theory. The crucial point is that the thermal mass increases with temperature as $T^2$ and this can restore the symmetry of the vacuum.

Consequently, there exists a critical temperature, $T_c$, beyond which the origin becomes the global minimum of the effective potential, i.e.,
\begin{eqnarray}
\langle\phi\rangle &=& 0\,,\qquad\,\,\, T>T_c\\ \nonumber
\langle\phi\rangle &=& v_\phi\,,\qquad T < T_c
\end{eqnarray}
The precise critical temperature depends on the finite-temperature
corrections to the scalar potential and therefore on the microscopic
particle content and couplings. A dedicated calculation of the
finite-temperature effective potential, including the nucleation and
completion of the transition, is beyond the scope of this work. We
therefore adopt a phenomenological approach and take $T_c\sim v_\phi $ for our choices of $m_\phi \sim \mathcal{O}(10\,{\rm MeV})$ as a representative benchmark, following the treatment of~\cite{Davoudiasl:2005fd}.

As the PNS cools through neutrino diffusion, the thermal correction gradually decreases. Once the temperature approaches $T_c$, the effective potential begins to recover the broken minimum characteristic of the zero-temperature theory. The subsequent evolution depends sensitively on the structure of the effective potential. If the transition between the symmetric and broken minima occurs continuously, the scalar expectation value evolves smoothly with temperature. In contrast, if the finite-temperature effective potential develops two locally stable minima separated by an energy barrier, the system undergoes a first-order phase transition.

The appearance of this barrier is a well-known consequence of finite-temperature field theory and may arise from thermal cubic terms, loop corrections, or additional interactions present in the scalar sector. Since our objective is to develop a model-independent phenomenological framework, we do not specify the microscopic origin of the barrier. Instead, we simply assume that the finite-temperature effective potential admits a first-order phase transition at a critical temperature relevant to the cooling evolution of the PNS. We also remain agnostic about nucleation dynamics, since these details, while important, are irrelevant for the neutrino signatures we are interested in. For a recent discussion on this matter, see~\cite{Bleau:2026ala}.

The transition is accompanied by the release of the vacuum-energy difference between the false and true vacua,
\begin{equation}
\Delta V
=V_{\rm false}- V_{\rm true}.
\end{equation}
Typically, a fraction of this vacuum energy can be stored in coherent oscillations of the scalar field. The remainder may be transferred to thermal particles, hydrodynamic motion of the surrounding medium, or other microscopic degrees of freedom depending on the underlying model. Since these processes are highly model dependent, we consider a phenomenological efficiency factor $f_{\rm PT}$ describing the fraction of vacuum energy converted into coherent oscillations of the scalar field about the true vacuum. The total energy stored in the condensate is therefore parameterised as
\begin{equation}
E_{\rm cond} = f_{\rm PT}\, \Delta V\, V_{\rm core},
\end{equation}
where $V_{\rm core}$ denotes the volume of the region undergoing the phase transition.

It is important to emphasise that these coherent oscillations are fundamentally different from the scalar particles populated in the plasma through interactions with neutrinos, as described in the previous section. The condensate is described by a single classical field configuration. Consequently, its decay produces a neutrino signal that is temporally and spectrally distinct from that arising from scalar particles. It is also important to highlight that the decay of the scalar field into neutrinos is independent of the mechanism of the phase transition and does not rely on the formation of bubbles. This is an additional ingredient which can lead to the formation of gravitational waves, complementing the neutrino signal.

The picture developed here naturally leads to two independent sources of neutrinos. The first consists of \emph{prompt} neutrinos produced through the decay of thermal scalar particles after they decouple from the PNS. The second consists of \emph{delayed} neutrinos produced through the gradual decay of the coherent scalar condensate generated by the first-order phase transition. Since both components originate from the same scalar field but correspond to different physical states, their spectra, temporal evolution, and relative normalisation are generally independent observables.

In the following section we derive the neutrino spectra associated with each of these emission channels and determine how they modify the observable neutrino signal from an individual SN.

\section{Impact of scalar induced FOPT on the neutrino spectrum}
\label{sec:Fluxes}
As discussed, the production of a thermal population of a $\mathcal{O}(10-100)$ MeV scalar and a possible vacuum symmetry restoration due to thermal corrections can eventually lead to prompt and delayed neutrino signals. In this section, we discuss the spectral identities of those two neutrino signals, emerging from the scalar decay.
\subsection{Prompt decay spectrum}
The prompt component originates from the thermal scalar particles that remain in equilibrium with the neutrino plasma throughout the symmetry-restored phase. 
The scalar population subsequently decouples at the scalar
transport sphere, after which the distribution is frozen out and the
scalars free stream out of the PNS.
As a result, the scalar mass transitions from its thermal value to the zero-temperature vacuum mass $m_\phi$. During free-streaming, the scalar can promptly decay into neutrinos.

We denote the three predominantly active light mass-eigenstates by $\nu_i, \,(i=1,2,3)$ and the predominantly sterile heavy state as $\nu_h$. For small active-sterile mixing angle $\theta \simeq y v_\phi /m_R$, the light mass states overlaps mostly with $\nu_L$, and similarly $\nu_h$ with $\nu_R$. We focus on the parameter region, $m_\phi < m_R \simeq m_h$, for which decays of $\phi$ into the heavy state $(\nu_h)$ are kinematically forbidden. The scalar therefore decays exclusively into the light neutrino mass-eigenstates $(\nu_i)$, and the resulting neutrinos inherit the Bose-Einstein distribution of their parent bosons, making them spectrally distinct from the primary thermal neutrinos.

Since both the thermal neutrinos and bosons were in equilibrium inside the neutrinosphere, the total energy of the CCSN is carried away by all the particle species. The total available relativistic degrees of freedom for the neutrino, anti-neutrino and the boson is $(6\times7/8)+1=50/8$~\cite{Davoudiasl:2005fd} \footnote{Note that this is an approximate calculation, which neglects the chemical potential of the neutrinos. A more rigorous calculation can improve this number, however, we checked that this does not change the final results significantly}.
As a benchmark estimate, we distribute the radiated energy among the
relativistic degrees of freedom in equilibrium, and hence the fraction $8/50\simeq0.16$ of it is carried away by the scalar $\phi$. Consequently, the number spectrum of $\phi$ follows a Bose-Einstein distribution normalised to carry $\sim16\%$ of the total radiated energy:
\begin{equation}
    \frac{dN_{\phi}}{dE_{\phi}}\simeq0.16\times E^{\mathrm{tot}}\frac{15}{T^4\pi^4}\frac{E_{\phi}^2}{e^{E_{\phi}/T}-1}\;.
\end{equation}

For a scalar decaying at rest, the daughter neutrinos each possess energy
$E_\nu=m_\phi/2$, up to negligible corrections from the neutrino masses. In practice, however, the thermal scalar particles possess non-zero momenta inherited from the Bose-Einstein distribution. Consequently, the Lorentz boost of the parent scalar broadens the neutrino energy distribution. The observed prompt neutrino spectrum therefore results from a convolution of the thermal scalar distribution with the two-body decay kinematics. As the scalar decays through the same coupling, which generates the neutrino mass, the number spectrum of the individual neutrino mass eigenstate is expected to be proportional to $m_i^2$ and is given by:
\begin{equation}
     F^{\mathrm{pr}}_{\nu_i}(E_{\nu_i})=\frac{dN_{\nu_i}}{dE_{\nu_i}}\simeq \frac{m_i^2}{2\sum m_k^2}\times0.16\times E_{\mathrm{tot}}\times2\int_{E_{\mathrm{min}}}^\infty dE_{\phi}\frac{1}{E_{\phi}}\frac{15}{T^4\pi^4}\frac{E_{\phi}^2}{e^{E_{\phi}/T}-1}\;.
     \label{eq: prompt_flux}
\end{equation}
Here, $m_i^2/\left(2\sum m_k^2\right)$ is the branching ratio for each  mass-eigenstate $\nu_i$, assuming the scalar decaying to neutrinos and antineutrinos equiproportionally.
The factor of two accounts for the two neutrinos produced in every scalar decay. Here, $E_{\mathrm{min}}$ is the minimum energy of the scalar to produce a neutrino with energy $E_{\nu}$.

After production, the mass-eigenstate flux propagates through the SN to the Earth. Their observable flavour composition is obtained by projecting the mass-eigenstate flux through the PMNS matrix into the corresponding flavour states. The flavour evolution can modify the relative flavour
composition, but does not alter the energy spectrum generated by the
scalar decay.

\subsection{Delayed decay spectrum}
The delayed neutrino component originates from the coherent scalar condensate formed during the first-order phase transition. 
Assuming that the phase transition primarily produces the zero-momentum scalar condensate, the resulting scalar configuration can be treated
as a coherent classical field oscillating about the true vacuum.
The corresponding number of scalar quanta contained within the condensate may therefore be approximated by
\begin{equation}
N_\phi^{\rm cond}=\frac{E_{\rm cond}}{m_\phi},
\end{equation}
where the small kinetic energy of the oscillating field has been neglected. For our analysis, we approximate the released energy is dominated by the quartic terms, hence $E_{\rm cond}\approx v_{\phi}^4f_{\rm PT} V_{\rm core}$.

The condensate decays to neutrino mass-eigenstates through the same interaction channel as the prompt decay scenario. Unlike the prompt component, whose spectral shape is determined by the Bose-Einstein distribution of the parent scalars, the delayed spectrum is determined primarily by the kinematics of condensate decay. Since the coherent field consists predominantly of non-relativistic scalar quanta, each decay produces neutrinos with energies close to $E_\nu\simeq m_\phi/2.$

This delayed decay neutrino spectrum of mass eigenstate $\nu_i$ can be written as
\begin{equation}
    F^{\mathrm{del}}_{\nu_i}(E_{\nu_i})=\frac{dN_{\nu_i}}{dE_{\nu_i}}= \frac{m_i^2}{2\sum m_k^2}\times2N^{\rm cond}_{\phi}\delta\left(E_{\nu_i}-\frac{m_{\phi}}{2}\right)\,.
    \label{eq:delayed_flux}
\end{equation}
Once produced, the neutrinos propagate through the surrounding PNS medium. For these neutrinos to provide an additional
observable component of the SN neutrino flux, their mean free path must
be sufficiently large that a significant fraction of them escape
without being reprocessed by the dense medium. In our phenomenological
treatment, we assume that the delayed neutrinos are produced in a
region sufficiently close to the neutrinosphere such that
they can free-stream out without significant trapping.

In a realistic SN environment, several effects are expected to broaden this monochromatic spectrum. Spatial variations of the condensate due to residual bulk motion of the scalar field, inhomogeneities generated during the phase transition, and the finite momentum dispersion of the oscillating field all contribute to a finite spectral width. However, for simplicity, we neglect this contribution and work with a monochromatic spectrum for our analysis.

\section{The diffuse supernova neutrino spectrum}
\label{sec:DSNB}
The previous discussion focuses on the dynamics of a single CCSN. However, since the time of star-formation, there have been countless CCSN in the history of the universe. Each CCSN produces a huge flux of neutrinos, originating from weak interactions inside the PNS. 
The DSNB is obtained by summing the contributions from all past CCSN since star-formation. These MeV neutrinos freely propagate over cosmological distances while their energies are redshifted by the expansion of the Universe. Consequently, the observed DSNB provides a unique probe of both the cosmic SN population and any new neutrino production mechanisms operating inside an individual SN.

The DSNB for individual flavour of neutrino can be computed as
\begin{equation}
    \Phi^{\rm conv}_0(\nu_\beta, E_{\nu_\beta}^0) =\int_0^{z_{\mathrm{max}}}\frac{dz}{H(z)}R_{\mathrm{CCSN}}(z)F^{\rm conv}_{\nu_\beta} (E_{\nu_\beta})\,,
    \label{eq: DSNB_std}
\end{equation}
where $F^{\rm conv}_{\nu_\beta}(E_{\nu_\beta})$ is the conventional neutrino spectrum at the source with energy $E_{\nu_\beta}=E_{\nu_\beta}^0(1+z)$ and $E_{\nu_\beta}^0$ is the redshifted energy observed at the Earth. The flux determination relies on a knowledge of the co-moving star formation rate (SFR), which is given by~\cite{Horiuchi:2008jz}:
\begin{equation}
    \dot{\rho}_*(z)=\dot{\rho}_0\left[(1+z)^{-10a}+\left(\frac{1+z}{B}\right)^{-10b}+\left(\frac{1+z}{C}\right)^{-10\gamma}\right]^{-1/10}\,.
\end{equation}
Here, $\dot{\rho}_0$ is an overall normalisation factor and $a$, $b$, $\gamma$ are dimensionless exponents. $B$ and $C$ are parameters related to redshift breaks at $z_1=1$ and $z_2=4$, which are given by:
\begin{equation}
    B=(1+z_1)^{1-\frac{a}{b}}
\end{equation}
\begin{equation}
    C=(1+z_1)^{\frac{b-a}{\gamma}}(1+z_2)^{1-\frac{b}{\gamma}}\,.
\end{equation}
Following~\cite{Horiuchi:2008jz}, we take $\dot{\rho}_0=0.0178_{-0.0036}^{+0.0035}~\mathrm{M}_\odot~\mathrm{yr}^{-1}~\mathrm{Mpc}^{-3}$, $a=3.4\pm 0.2$, $b=-0.3\pm0.2$ and $\gamma=-3.5\pm1$. Furthermore, $R_{\mathrm{CCSN}}$ is taken to be related to SFR as,
\begin{equation}
    R_{\mathrm{CCSN}}(z)=\frac{\dot{\rho}_*(z)}{143M_\odot}\,.
\end{equation}
Finally, $H(z)$ is the Hubble parameter determined from the Friedmann equation with $H_0=67.36~\mathrm{km}~\mathrm{s}^{-1}~\mathrm{Mpc}^{-1}$. In our work, we have considered the maximum redshift for significant star formation to be $z_{\mathrm{max}}=5$.

The neutrino flux $F^{\rm conv}_{\nu_\beta}(E_{\nu_\beta})$ has contributions from both CCSNe and black-hole-forming (BHF) failed supernovae such that:
\begin{equation}
   F^{\rm conv}_{\nu_\beta}(E_{\nu_\beta})=f_{\mathrm{SN}}F^{\mathrm{SN}}_{\nu_\beta}(E_{\nu_\beta})+f_{\mathrm{BH}}F^{\mathrm{BH}}_{\nu_\beta}(E_{\nu_\beta}).
\end{equation}
Here, $f_{\mathrm{SN}}$ and $f_{\mathrm{BH}}$ represent the fraction of events leading to CCSNe and BHF, respectively. We have chosen a benchmark value of $f_{\mathrm{BH}}=20\%$, which lies between the estimate of $18\%-42\%$. The quantities $F^{\mathrm{SN}}_{\nu_\beta}(E_{\nu_\beta})$ and $F^{\mathrm{BH}}_{\nu_\beta}(E_{\nu_\beta})$ are the time integrated neutrino spectra from CCSNe and BHF-SNe. We have taken the standard ``alpha-fit'' neutrino spectrum 
given by:
\begin{equation}
    F_{\nu_\beta}^{\mathrm{SN/BH}}(E_{\nu_\beta})=\frac{0.84\times E_{\nu_\beta}^{\mathrm{tot}}}{\langle E_{\nu_\beta}\rangle^2}\frac{(\alpha_{\nu_\beta}+1)^{(\alpha_{\nu_\beta}+1)}}{\Gamma(\alpha_{\nu_\beta}+1)}\left(\frac{E_{\nu_\beta}}{\langle E_{\nu_\beta}\rangle}\right)^{\alpha_{\nu_\beta}}\mathrm{exp}\left[-(\alpha_{\nu_\beta}+1)\left(\frac{E_{\nu_\beta}}{\langle E_{\nu_\beta}\rangle}\right)\right],
\end{equation}
where, $\langle E_{\nu_\beta}\rangle$~($\nu_\beta\in \nu_e,\bar{\nu}_e, \nu_x$) is the average neutrino or antineutrino energy of flavour $\beta$, $\Gamma$ is the Euler gamma function and $\alpha$ is the dimensionless pinching parameter. As discussed before, the total luminosity gets equipartitioned into all the available degrees of freedom, including the scalar, and hence, the fraction $42/50\simeq0.84$ of it is carried away by the neutrinos. For this work, as a representative case, we have considered the time integrated spectrum from a CCSNe and a BHF-SNe with mass 27 ${\rm M}_\odot$ and 40 ${\rm M}_\odot$, respectively, from the Garching group simulation results~\cite{Kresse:2020nto}.

In the standard picture, the DSNB is entirely determined by the conventional neutrino emission associated with gravitational core collapse. In the present scenario, however, every SN additionally contributes prompt and delayed neutrino components originating from the scalar sector. Since these new emission channels are generated within the SN itself, they enter the cosmological calculation on exactly the same footing as the standard neutrino signal.

The standard flux $\Phi^{\rm conv}$ and the prompt flux $\Phi^{\rm pr}$ receive distinct contributions from CCSN and BHF-SN because of differences in the parameters. We neglect the delayed component from BHF events, assuming that the
formation of the BH terminates the PNS evolution before the
phase transition and subsequent condensate decay can occur. Hence, the two distinct diffuse  neutrino fluxes for the prompt and delayed spectra are given, respectively, by:
\begin{align}
\Phi_0^{\mathrm{pr}}(\nu_i,E^0_{\nu_i}) &= \int_0^{z_{\mathrm{max}}}\frac{dz}{H(z)}R_{\mathrm{CCSN}}(z)\left[f_{\mathrm{SN}}F^{\mathrm{pr(SN)}}_{\nu_i}(E_{\nu_i})+f_{\mathrm{BH}}F^{\mathrm{pr(BH)}}_{\nu_i}(E_{\nu_i})\right]
\label{eq: DSNB_prompt}\,,\\
\Phi_0^{\mathrm{del}}(\nu_i,E^0_{\nu_i}) &=\int_0^{z_{\mathrm{max}}}\frac{dz}{H(z)}R_{\mathrm{CCSN}}(z) f_{\mathrm{SN}}F^{\mathrm{del}}_{\nu_i}(E_{\nu_i})
\label{eq: DSNB_delayed}\,.
\end{align}

\begin{figure}[tbp]
    \centering
    \includegraphics[width=0.48\linewidth]{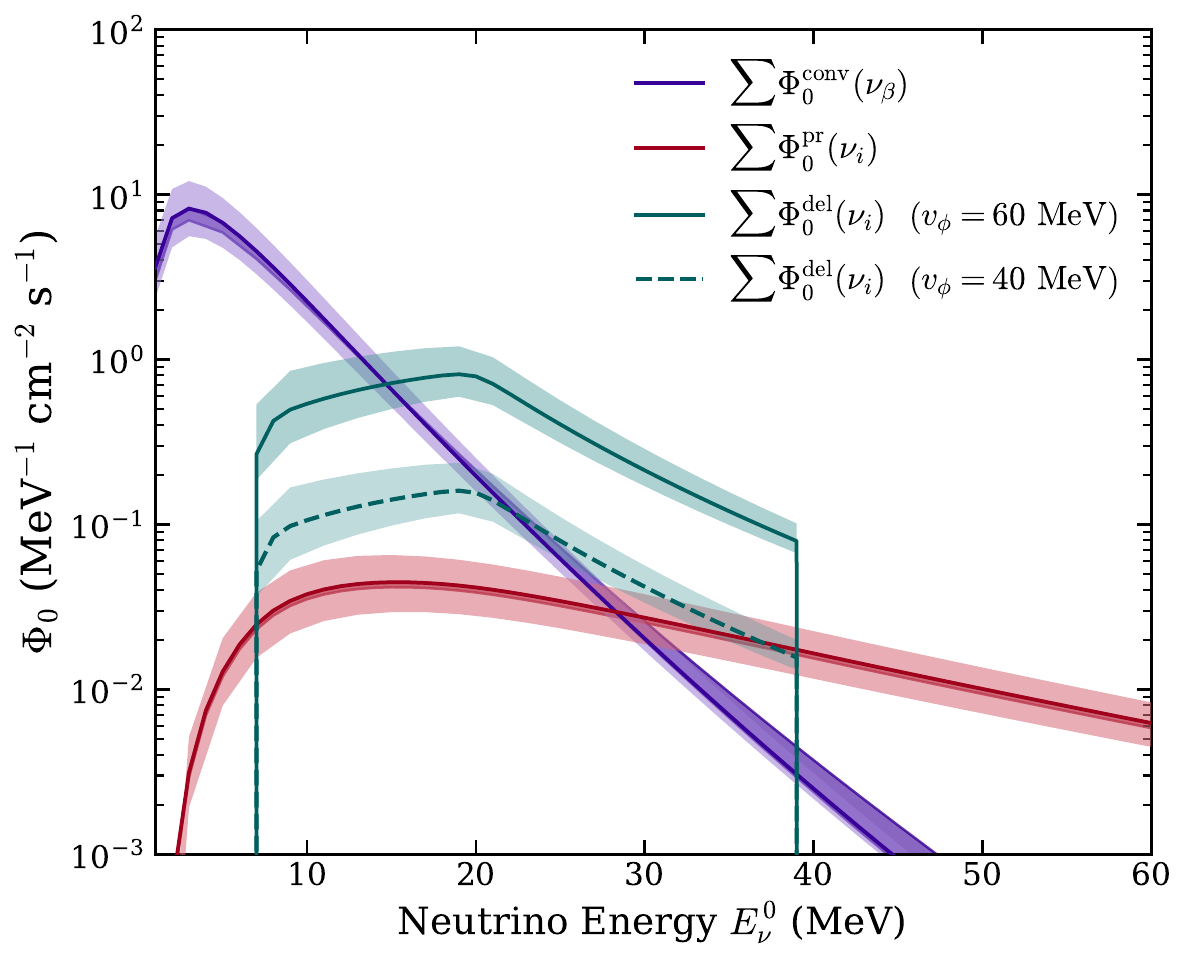}
    \includegraphics[width=0.48\linewidth]{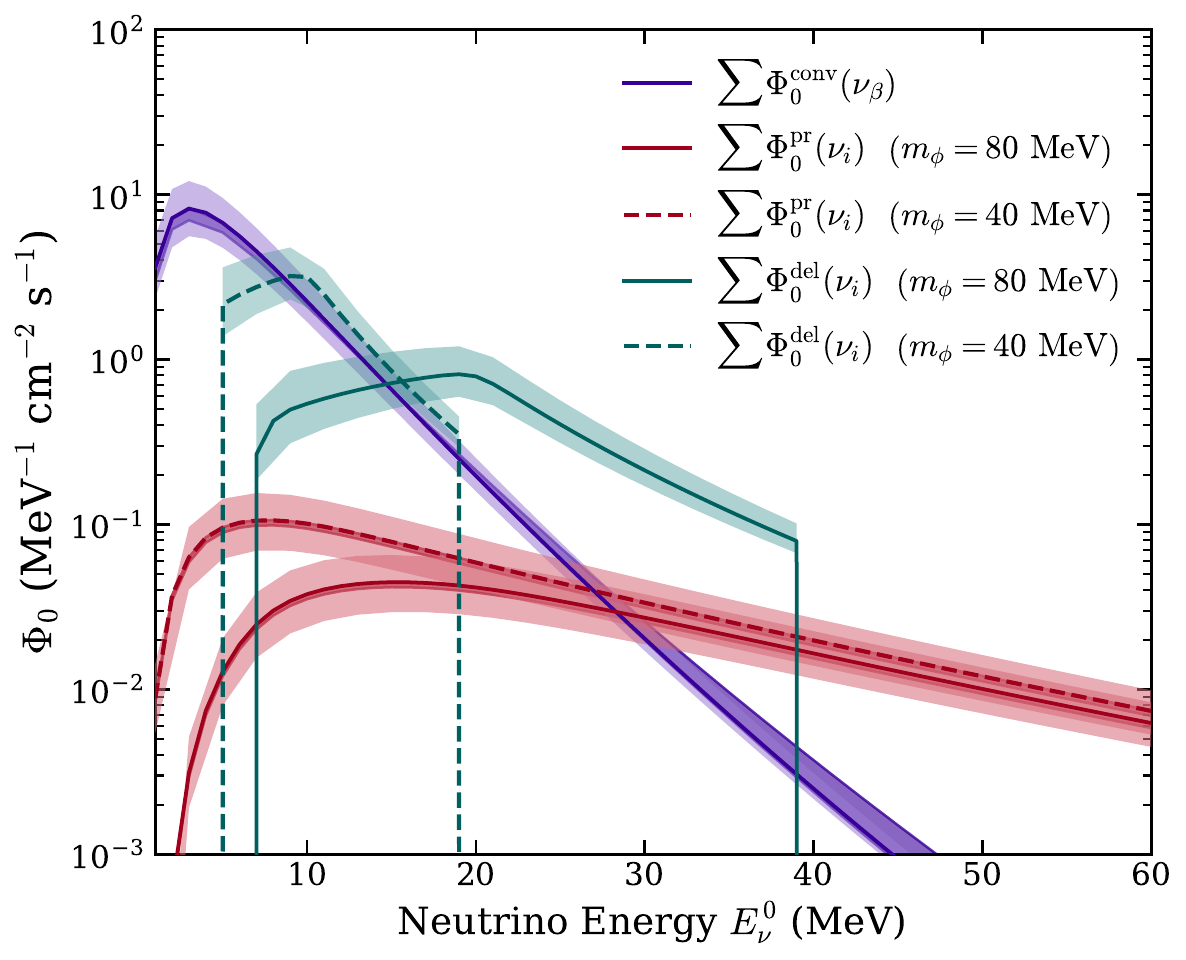}
\caption{Standard and new physics contribution to the total (summed over all states) DSNB flux. 
Purple line shows standard DSNB flux, while coral~(teal) lines correspond to the prompt decay (delayed decay) contributions (see sec.~\ref{sec:Fluxes} for details) computed using best-fit values of the SFR parameters and $f_{\rm BH} = 0.2$. Light colour bands shows the uncertainties in the DSNB flux due to the SFR uncertainties while the dark colour band shows the same due to the black-hole formation rate uncertainties, $f_{\rm BH}\in [0.18,0.42]$~\cite{Kresse:2020nto}. The left panel shows the results for $v_\phi = (40$ MeV, $60$ MeV), with $m_\phi = 80$ MeV. The right panel shows the results for $m_\phi = (40$ MeV, $80$ MeV), with $v_\phi = 60$ MeV.} 
    \label{fig:DSNB_flux}
\end{figure}

In Fig.\,\ref{fig:DSNB_flux}, we depict the different components of the total DSNB flux, defined in Eqs.\,\ref{eq: DSNB_std},~\ref{eq: DSNB_prompt},~\ref{eq: DSNB_delayed}, as functions of the neutrino energy. Benchmark values of $\mp$, $v_\phi$, and $T$ are considered such that the scalar undergoes a phase transition within the CCSN.

The new-physics contributions from prompt and delayed scalar decays are shown by the coral and teal bands respectively, while the standard DSNB flux is represented by the purple band. At low neutrino energies, the standard DSNB flux dominates over the new-physics contributions for the chosen scalar parameters. As the neutrino energy increases, the standard DSNB flux decreases, allowing the scalar-induced contributions to become increasingly important. The delayed $\phi$-decay component is confined to the energy range $\E \in [\mp/\left(2 (1+z_{\rm max})\right),\mp/2]$, which can be explained from Eq.\,\ref{eq:delayed_flux}. In contrast, neutrinos produced through prompt $\phi$ decays inherit the thermal Bose-Einstein distribution of the scalar population emitted from the SN core, resulting in a broader spectral contribution.

The left panel illustrates the dependence of the total DSNB flux on $v_\phi$. We present results for two benchmark values, $v_\phi = 40$ MeV and $60$ MeV, while keeping $\mp = 80$ MeV fixed. We consider a fixed efficiency factor $f_{\rm PT} = 0.75$ throughout this work.
Since the energy released during the phase transition, $E_{\rm cond}$, increases rapidly with $v_\phi$, the delayed neutrino flux correspondingly grows with $v_\phi$.  For sufficiently large values, $v_\phi \gtrsim 35\,$MeV, the delayed contribution can become larger than the standard DSNB flux. In contrast, the prompt contribution is independent of $v_\phi$, as its normalisation is determined by the thermal scalar population rather than by the energy released during the phase transition.

The right panel shows the dependence on the scalar mass, $\mp$, presenting $\mp = 40\,$MeV and $80\,$MeV while keeping $v_\phi=60\,$MeV. We find that, unlike $v_\phi$, changing $\mp$ affects both prompt neutrino and delayed neutrino contributions to the DSNB flux. As $\mp$ increases, the prompt neutrino spectrum shifts toward larger energy, since the lower integration limit in Eq.\,\ref{eq: prompt_flux} increases with $\mp$. A similar shift occurs for the delayed component: the characteristic energy range of the delayed neutrino spectrum moves toward higher energies as $\mp$ increases. At the same time, the overall delayed flux decreases with increasing $\mp$ as the number of scalars produced due to the phase transition decreases.

Note that refs.~\cite{Akita:2022etk, Fiorillo:2022cdq} investigated a related scenario in which the emission of a Majoron-like boson from the SN core and its subsequent decay into neutrinos modifies the neutrino spectra and hence the DSNB. However, these studies primarily considered the small-coupling regime, $y\, m_\phi\lesssim 10^{-7}\,{\rm MeV}$~(for a 100 MeV boson), where Majoron predominantly decays outside the SN core. In contrast, our study considers a large-coupling regime ($y\, \mp \gtrsim 10^{-6}$ MeV) in which a significant fraction of the scalars remain trapped inside the SN core and can subsequently participate in the phase transition. Nevertheless, the prompt-decay component in our scenario is closely related to the mechanism considered in refs.~\cite{Akita:2022etk,Fiorillo:2022cdq}, and consequently exhibits similar qualitative features in its spectral shape, as shown in Fig.\,\ref{fig:DSNB_flux}.

\section{DSNB signal analysis at neutrino experiments}
\label{sec:Events_analysis}
We now proceed to calculate the net expected neutrino signal events from the DSNB at the currently operating Super-K detector, as well as the next-generation neutrino experiments DUNE and Hyper-K. The differential event spectrum as a function of reconstructed neutrino energy is given by
\begin{align}
  \frac{dN}{dE_{\nu}^{\rm rec}} = N_t T \epsilon \int_{E^{\rm tr}_{\rm min}}^{E^{\rm tr}_{\rm max}} dE^{tr}\, \Phi(E^{\rm tr}) \,\sigma(E^{\rm tr})\,W(E^{\rm tr},E^{\rm rec})\,,
  \label{eq:event_rate}
\end{align}
where $N_t$ is the number of target particles, $T$ is the detector exposure time, and $\epsilon$ denotes the detection efficiency. The quantities $E^{\rm tr}$ and $E^{\rm rec}$ are the true and reconstructed neutrino energies, respectively, while $\sigma(E^{\rm tr})$ is the relevant neutrino interaction cross section and $W(E^{\rm tr},E^{\rm rec})$ is the detector energy-resolution function.

Recently, the Super-K collaboration reported its results 
from 956.2 days of data taking using the 22.5 kton fiducial-volume Gd-doped water-Cherenkov detector 
~\cite{Super-Kamiokande:2025sxh}. The DSNB is detected primarily through inverse beta decay (IBD),$\bar{\nu}_e + p\rightarrow e^+ + n$. The addition of gadolinium significantly improves neutron tagging and consequently enhances the DSNB sensitivity. We adopt the same exposure and take a detection efficiency of $\epsilon=67\%$.
The IBD cross section is taken from ref.~\cite{Strumia:2003zx}, while the detector response is modeled with a Gaussian energy-resolution function, with resolution $\sigma_E = 0.6\times\sqrt{E^{\rm rec}}$ following~\cite{DeGouvea:2020ang}.

Hyper-K, the next-generation water-Cherenkov detector, will provide a substantially larger target volume. We consider a 187 kton fiducial mass with gadolinium doping and assume the same detection efficiency and energy-resolution function as for Super-K. We take a total exposure of 10 years. For both Super-K and Hyper-K, the principal backgrounds arise from atmospheric $\nu_e$ and $\bar{\nu}_e$, as well as events associated with invisible-muon decays. At lower energies, cosmic-ray muon spallation provides an additional background, including the production of radioactive isotopes such as $^{9}\mathrm{Li}$.

DUNE provides complementary sensitivity to the $\nu_e$ component of the DSNB through charged-current interactions in liquid argon, $\nu_e+ {\rm Ar} \rightarrow {\rm K}^\ast + e^-$. We consider a 40 kton liquid-argon detector with a detection efficiency of $\epsilon=86\%$. The energy resolution is modeled by a Gaussian with $\sigma_E = 0.11\sqrt{E^{\rm rec}}+ 0.02\times E^{\rm rec}$. Similar to Hyper-K, we assume a 10-year exposure. The dominant backgrounds are solar neutrinos at $E\lesssim16$ MeV and atmospheric $\nu_e$ at higher energies, $E\gtrsim30$ MeV.


\begin{figure}[tbp]
    \centering
    \includegraphics[width=0.6\linewidth]{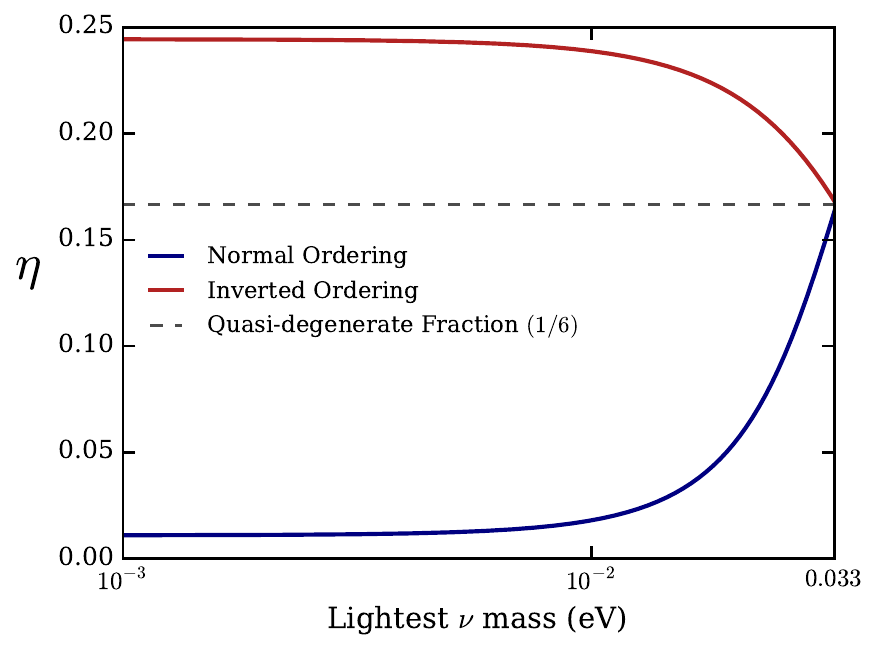}
    \caption{Prompt and delayed $\nu_e$~($\bar{\nu}_e$) flux fraction, $\eta$, as a function of the lightest neutrino mass for both NMO and IMO. The suppression in the case of NMO comes from the dominant contribution of $|U_{e3}|^2$, whereas in the case of IMO, it is dominated by both $|U_{e1}|^2$ and $|U_{e2}|^2$. In the limit $m_1\approx m_2\approx m_3\approx0.1~\mathrm{eV}/3$, the flux fraction converges to 1/6 in both the scenarios.}
    \label{fig: flux fraction}
\end{figure}

The flux entering Eq.\,\ref{eq:event_rate} is the corresponding $\nu_e$ or $\bar{\nu}_e$ flux at Earth. For the conventional DSNB flux, we know that the neutrinos decohere by the time they arrive at the Earth. Hence, the 
$\nu_e$~($\bar{\nu}_e$) flux at Earth, including neutrino mixing effect, is given by
\begin{align}
\Phi^{\mathrm{conv}}(\nu_e,E^0_{\nu_e}) = |U_{ei}|^2 \Phi_0^{\mathrm{conv}}(\nu_e,E^0_{\nu_e})+(1-|U_{ei}|^2) \Phi_0^{\mathrm{conv}}(\nu_x,E^0_{\nu_x})\,,
\label{eq:flux_std}
\end{align}
where  $U_{e i}$ denotes the PMNS mixing matrix elements. For normal mass ordering (NMO), $i=3$ for $\nu_e$ and $i=1$ for $\bar{\nu}_e$, while for inverted mass ordering (IMO), $i=2$ for $\nu_e$ and $i=3$ for $\bar{\nu}_e$.

The treatment of the scalar-induced flux differs from the standard DSNB because the scalar decays directly produce neutrino mass-eigenstates. Hence, the diffuse prompt and delayed fluxes of $\nu_e$ and $\bar{\nu}_e$ at Earth are consequently
\begin{align}
    \Phi^{\mathrm{pr}}(\nu_e,E^0_{\nu_e}) &= \sum_i|U_{ei}|^2\Phi_0^{\mathrm{pr}}(\nu_i,E^0_{\nu_i})\,,\,\,\,\,\,\label{eq:prompt2}\\
    \
    \Phi^{\mathrm{del}}(\nu_e,E^0_{\nu_e}) &= \sum_i|U_{ei}|^2\Phi_0^{\mathrm{del}}(\nu_i,E^0_{\nu_i})\,.
    \label{eq:delayed2}
\end{align}

Substituting Eq.~\ref{eq: DSNB_prompt} (\ref{eq: DSNB_delayed}) into Eq.~\ref{eq:prompt2} (\ref{eq:delayed2}), one finds that the prompt/delayed $\nu_e$ and $\bar{\nu}_e$ fluxes are proportional to the flux fraction factor $\eta$,
\begin{equation}
\eta=\left(\frac{1}{2\sum_k m_k^2}\right)\sum_i |U_{ei}|^2m_i^2=\frac{\Phi^{\mathrm{pr/del}}(\nu_e,E^0_{\nu_e})}{2\sum_i\Phi_0^{\mathrm{pr/del}}(\nu_i,E^0_{\nu_i})}\,,
    \label{eq: flux fraction}
\end{equation}
which can be defined as the ratio of the $\nu_e$~(or $\bar{\nu}_e$) component of the prompt/delayed spectrum at Earth with respect to the total spectrum. Since $\eta$ depends on the mixing parameters $|U_{ei}|^2$ and the mass-squared values $m_i^2$, it is sensitive to both the neutrino mass ordering and the mixing angles. Fig.\,\ref{fig: flux fraction} shows the prompt and delayed flux fractions of $\nu_e$ and $\bar{\nu}_e$ for both the normal mass ordering (NMO) and the inverted mass ordering (IMO).

For NMO, the relevant projection is dominated by the small factor $|U_{e3}|^2$, and hence, the contribution to the electron-flavour flux is strongly suppressed when the lightest neutrino mass is small. For IMO, the corresponding flux receives contributions from both $|U_{e1}|^2$ and $|U_{e2}|^2$, resulting in a substantially weaker suppression.  As the lightest neutrino mass increases, the spectrum approaches the quasi-degenerate (QD) regime, $m_1\sim m_2\sim m_3$, in which the mixing dependence simplifies and both scalar-induced flux fractions approach $\sim 1/6$.

\begin{figure}[tbp]
\centering
\begin{subfigure}{0.48\textwidth}
\centering
\includegraphics[width=\linewidth]{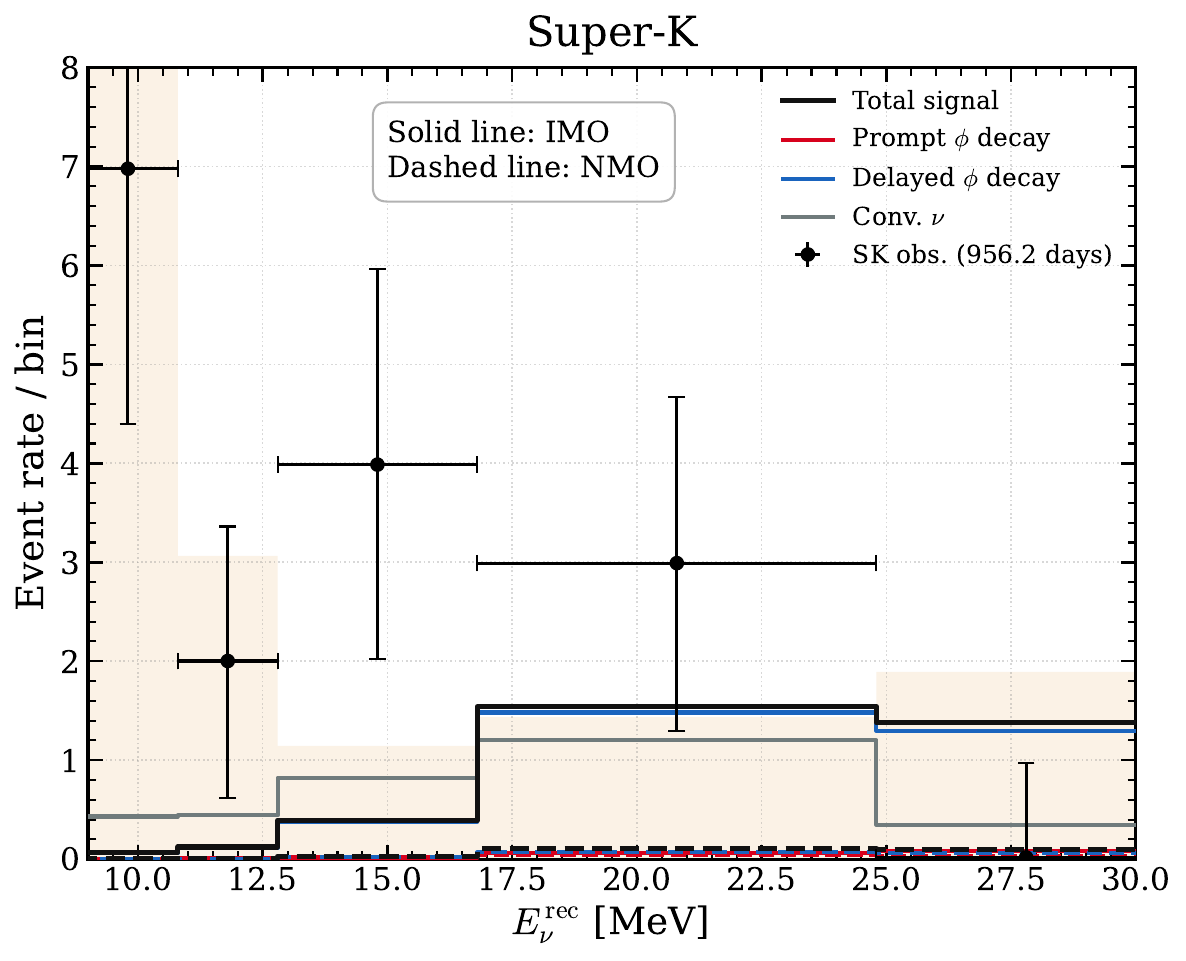}
\end{subfigure}
    \begin{subfigure}{0.48\textwidth}
        \centering
        \includegraphics[width=\linewidth]{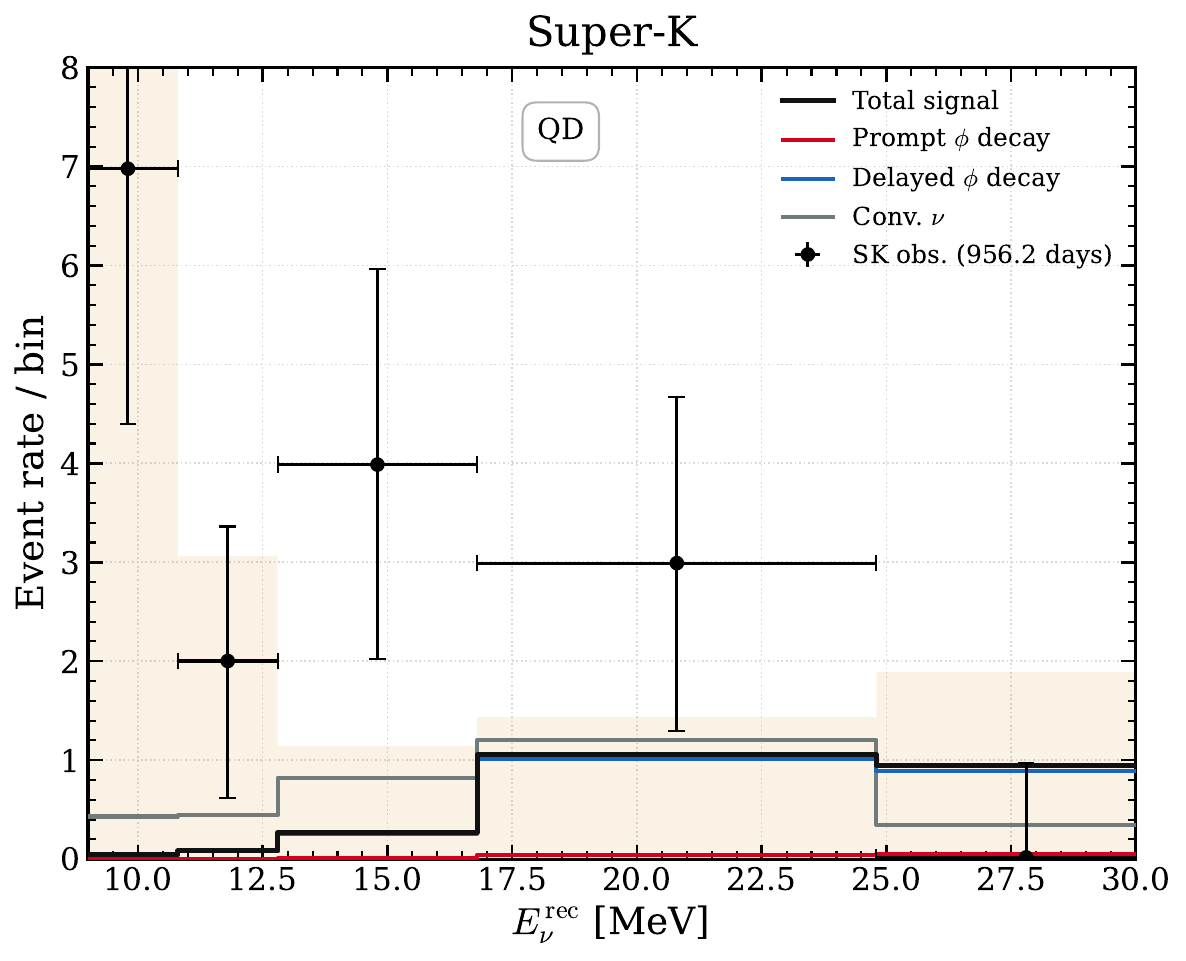}
    \end{subfigure}
        \caption{Net $\bar{\nu}_e$ event rate at Super-K with 956.2 days of exposure. The left panel presents the expected event rate for IMO  (solid lines) and NMO (dashed lines), while the right panel shows the QD scenario, $m_1\sim m_2\sim m_3$.
     Blue~(Red) line denotes the contribution from delayed~(prompt) scalar decay, calculated assuming $\mp= 80$ MeV and $v_\phi = 60$ MeV, while the black line shows the total neutrino events from the scalar decay. Data points show the observed event rates in Super-K with 956.2 days of operation~\cite{Super-Kamiokande:2025sxh}. The shaded regions represent the total backgrounds events from the atmospheric neutrinos, reactor neutrinos, and Li spallation.
     Information on the background event rate is taken from ref.~\cite{Super-Kamiokande:2025sxh}.}
    \label{fig:event_rate_SK}
\end{figure}

\begin{figure}[tbp]
\centering
\begin{subfigure}{0.48\textwidth}
\centering
\includegraphics[width=\linewidth]{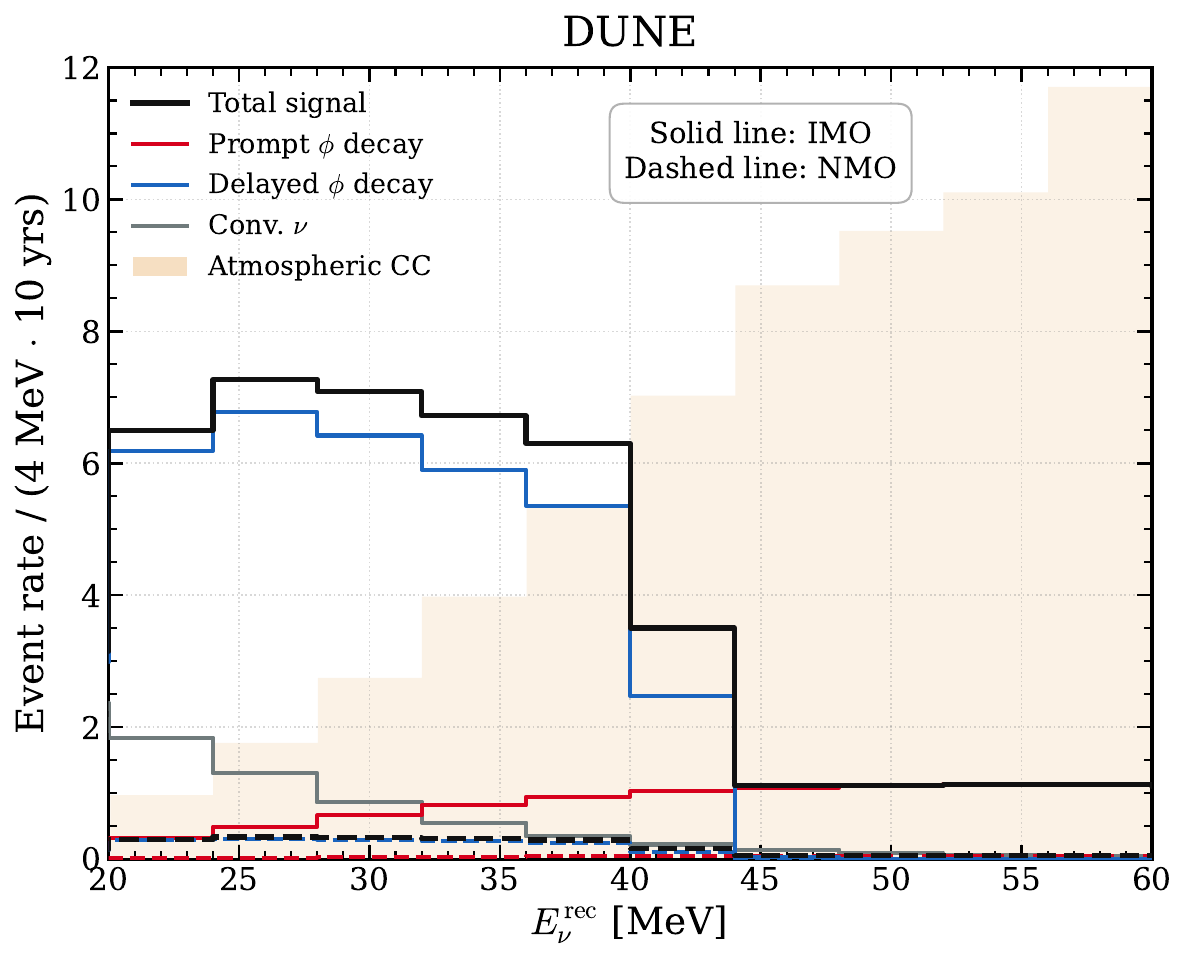}
\end{subfigure}
    \begin{subfigure}{0.48\textwidth}
        \centering
        \includegraphics[width=\linewidth]{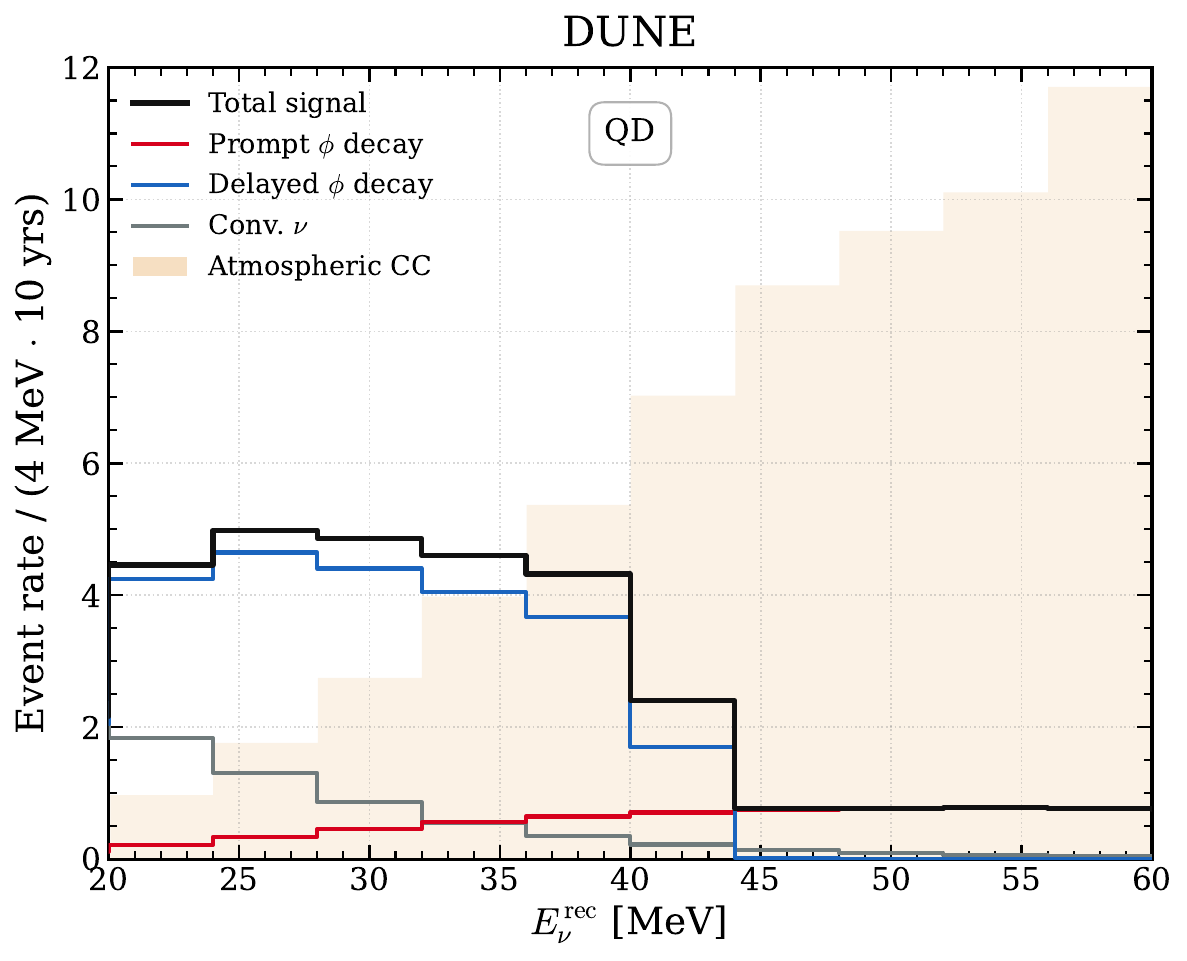}
    \end{subfigure}
    \vspace{0.5cm}
    \begin{subfigure}{0.48\textwidth}
        \centering
        \includegraphics[width=\linewidth]{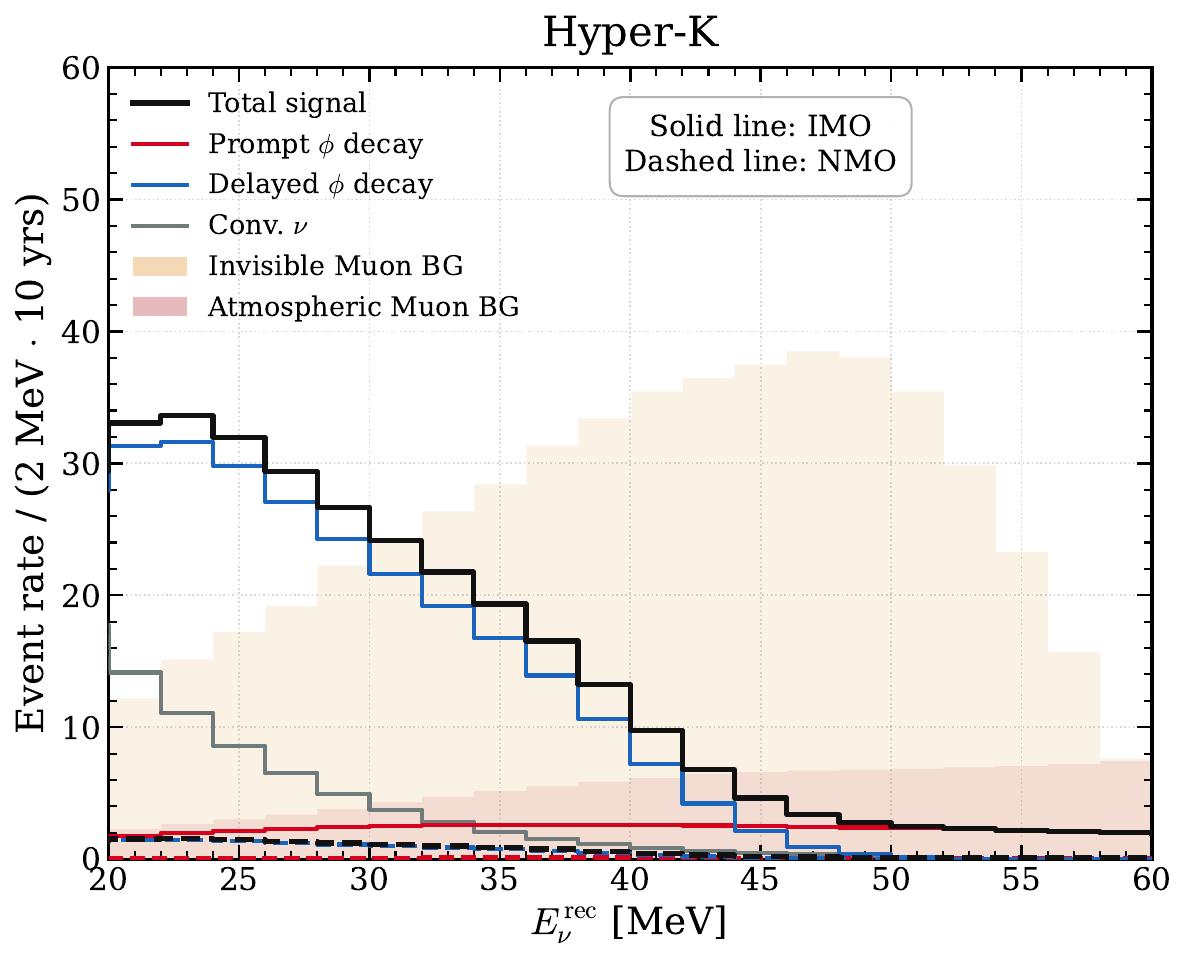}
    \end{subfigure}
    \begin{subfigure}{0.48\textwidth}
        \centering
        \includegraphics[width=\linewidth]{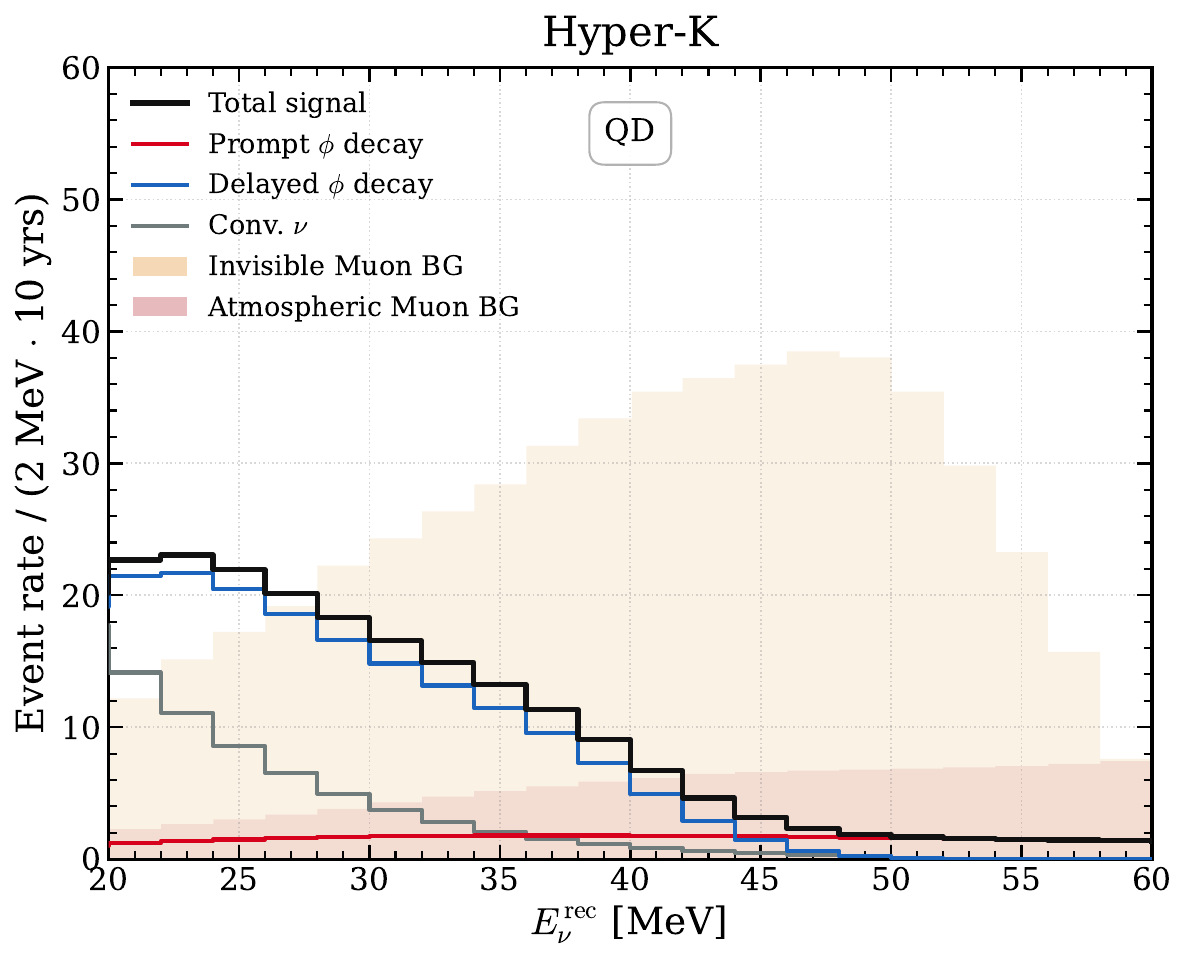}
    \end{subfigure}
    \caption{Expected $\nu_e$ and $\bar{\nu}_e$ event rate at DUNE (top panels) and Hyper-K (bottom panels). The left panel presents the expected event rate for two mass ordering cases, solid lines for IMO and dashed lines for NMO. The right panel shows the QD scenario, $m_1\sim m_2\sim m_3$.
    Blue~(Red) line denotes the contribution from delayed~(prompt) scalar decay, calculated assuming $\mp= 80$ MeV an $v_\phi = 60$ MeV, while the black line shows the total neutrino events from the scalar decay.  Information on the background event rate for DUNE and Hyper-K are taken from ref.~\cite{Akita:2022etk}.}
    \label{fig:event_rate}
\end{figure}
Fig.\,\ref{fig:event_rate_SK} shows expected DSNB $\bar{\nu}_e$ event rate at Super-K with 956.2 days of exposure, following a similar bin-structure for the reconstructed neutrino energy~\cite{Super-Kamiokande:2025sxh}. As a comparison, we also present the observed event rates in Super-K with 956.2 days of operation~\cite{Super-Kamiokande:2025sxh}. These event rates are computed for a benchmark value of $m_{\phi} = 80$ MeV, and $v_\phi = 60$ MeV, ensuring that $E_{\rm cond}$ is always less than SN total energy~$E^{\rm tot}_\nu$, assuming an average core radius of 25 km. The neutrino mixing parameters required to compute the electron neutrino fluxes are taken from ref.~\cite{Esteban:2024eli}.

We consider two limiting neutrino-mass scenarios: a hierarchical spectrum with negligible mass for the lightest neutrino, corresponding to $m_1 \approx 0$ for NMO, and $m_3 = 0$ for IMO (shown by the dashed and solid line, respectively, in left panels), and QD spectrum with $m_1\sim m_2 \sim m_3$ (shown in the right panel). For NMO case, neutrino signal from the prompt decay and delayed $\phi$ decay is negligible due to the $|U_{e3}|^2$ suppression. However, for the IMO case, new physics signal from the $\phi$-decay can be significantly enhanced due to the large value of the  flux fraction $\eta$, dominating over the standard DSNB signal in the last two bins. For QD scenario, signal rate has similar behavior as in IMO case, with a slightly smaller event rate due to a smaller $\eta$ value compared to the IMO scenario. Clearly, the existing data from Super-K can already be used to constrain the parameters of the scalar induced FOPT models.

Additionally, we also show the projected $\nu_e$ and $\bar{\nu}_e$ event rates at DUNE and Hyper-K for a 10-year exposure in Fig.\,\ref{fig:event_rate} (hierarchical scenario in the left panels, and QD scenario in the right panel). 
The top panel presents the results for a DUNE-like experiment, with a bin-width of 4 MeV. The gray lines represents the conventional DSNB event rate computed using the best-fit values of SFR parameters. 
As expected for the hierarchical NMO case, the scalar-induced $\nu_e$ signal at DUNE is strongly suppressed. In contrast, for IMO, the scalar contribution can exceed the conventional DSNB rate over a substantial part of the observable energy range. The lower-energy region is dominated by the delayed component, while the prompt contribution becomes increasingly important toward higher energies. At sufficiently high energies, however, the atmospheric charged-current background reduces the sensitivity to the scalar-induced signal. In the QD scenario, the mixing suppression is absent, resulting in a substantially larger event rate.

The bottom panels show the expected $\bar{\nu}_e$ event rate at the Hyper-K detector with 10 years of runtime with a bin width of 2 MeV. Expected event rate at Hyper-K shows similar features as in the case of DUNE. The larger fiducial volume of Hyper-K leads to a substantially larger number of expected events, while atmospheric neutrinos and invisible-muon decays constitute the dominant backgrounds.


\section{Sensitivity studies using Super-K, Hyper-K and DUNE}
\label{sec: sensitivity}
\begin{figure}[tbp]
    \centering
    \includegraphics[width=0.48\linewidth]{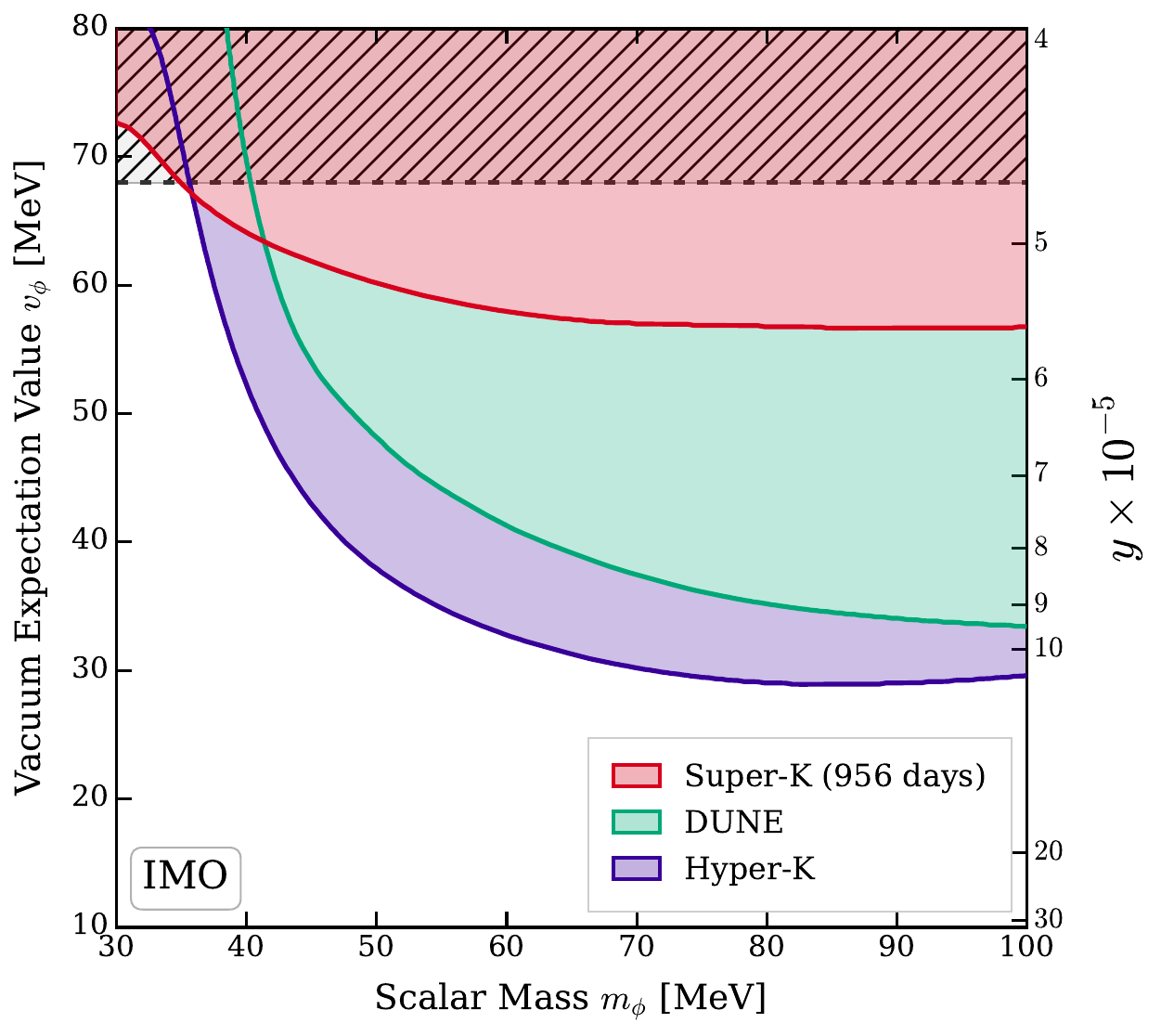}
    \includegraphics[width=0.48\linewidth]{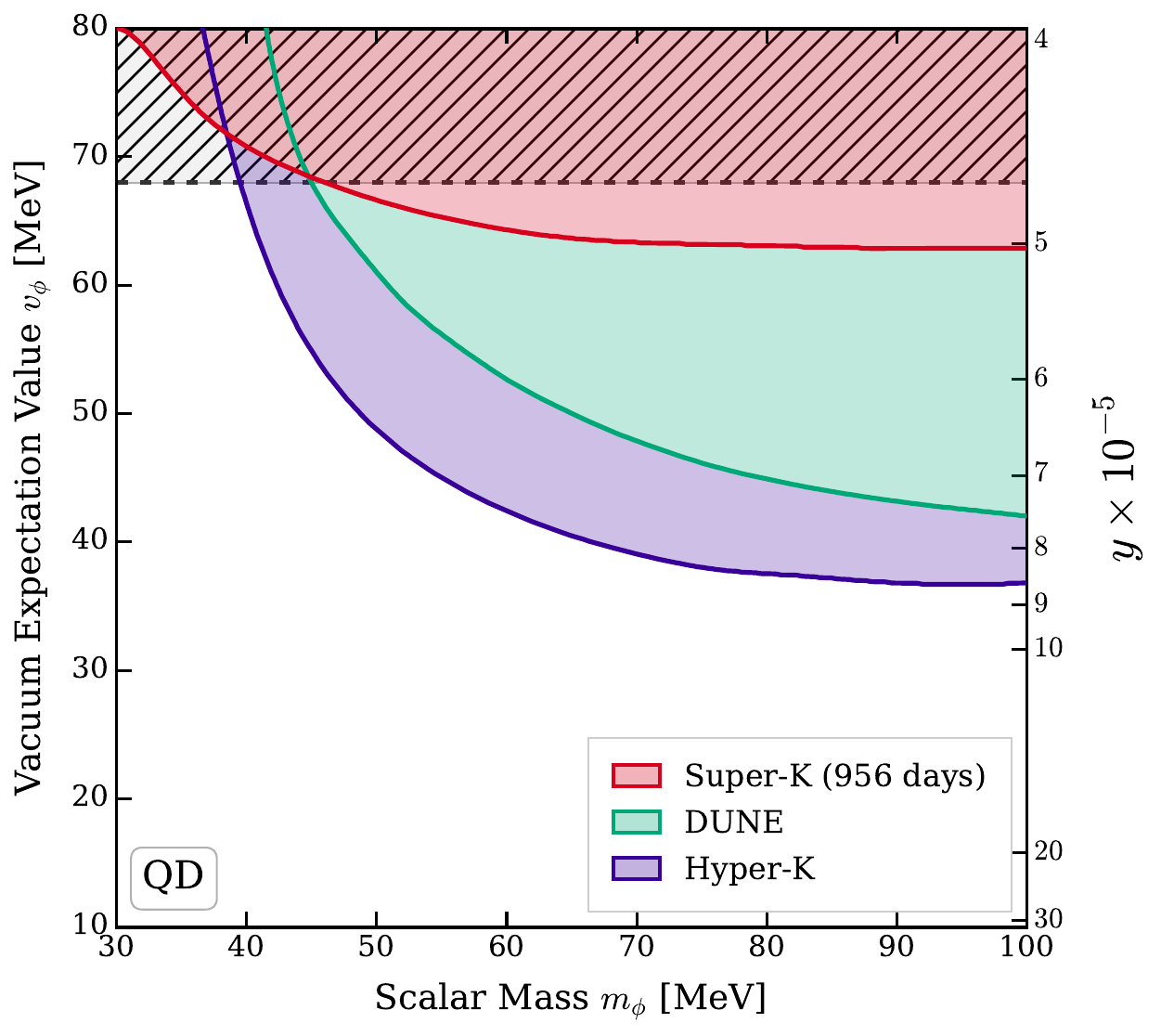}
    \caption{Excluded region in the ($m_\phi,v_\phi$) plane at 95\% confidence level. The red shaded region presents the constraints obtained from the data published by the Super-K collaboration~\cite{Super-Kamiokande:2025sxh}, while the teal and purple shaded region shows the projected constraints from DUNE and Hyper-K respectively. The top hatched exclusion contour comes from the fact that the amount of energy released in the FOPT has to be less than that of in CCSN. The corresponding Yukawa couplings $(y)$ are also depicted, which have been computed for $m_R=100$\,MeV. The left panel shows the result of IMO scenario, whereas the right panel shows the result for QD mass of neutrinos~($m_1\sim m_2\sim m_3$).}
    \label{fig:exclusion_plot}
\end{figure}
We now use the reported results from  Super-K and the future projections of DUNE and Hyper-K with 10 years of exposure to constrain the scalar parameter space $(\mp\,,v_\phi)$. We perform a binned Poisson-likelihood analysis, following the statistical treatment adopted in~\cite{Moller:2018kpn, Akita:2022etk},
\begin{align}
    \chi^2 = \min_{a_j,b_j}\left[-2\sum_i{\rm ln}\frac{L_{0,i}}{L_{1,i}}+\frac{a_j^2}{\sigma^2_{a_j}}+\frac{b_j^2}{\sigma^2_{b_j}}\right]\,,
 \end{align}
where likelihood functions $L_{x,i}$ are the probability of observing $k_i$ events at $i$-th bin with mean $\lambda_{x,i}$, defined in terms of Poisson distribution $ L_{x,i}=\left(\lambda_{x,i}^{k_i}\exp(-\lambda_{x,i})/k_i!\right)$. The nuisance parameters $a_j$ and $b_j$ accounts for the uncertainty in the measurements of signal and background, respectively. For our analysis, $\lambda_{0,i}$ and $\lambda_{1,i}$ are defined as follows
\begin{align}
\lambda_{0,i} &= \prod_{j}(1+b_j) N_{{\rm BG},i}\,,\\
\lambda_{1,i} & = \prod_j(1+ a_j) N_{{\rm sig},i} +\prod_{j}(1+b_j) N_{{\rm BG},i}\,,
\end{align}
where $N_{{\rm sig},i}$ and $N_{{\rm BG},i}$ denotes the number of signal events and background event rate at $i$-th bin. The product over $j$ accounts for the total effects of different systematic uncertainties in the signal and backgrounds.
The signal consists of the prompt and delayed scalar-decay contributions, while the background includes the standard DSNB flux together with atmospheric-neutrino and atmospheric-muon backgrounds. 

For Super-K, we use the same energy range and binning as the current DSNB analysis of~\cite{Super-Kamiokande:2025sxh}, covering an energy range (7.5, 30) MeV. We adopt $30\%$ uncertainty on the standard DSNB flux and a $25\%$ uncertainty on the atmospheric background. For Hyper-K,  we consider the 20-40 MeV range with 2 MeV bins and use the same systematic uncertainties as Super-K. For DUNE, we consider 20-60 MeV with 4 MeV bins. In addition to the DSNB-flux uncertainty, we include a $15\%$ uncertainty associated with the neutrino-argon cross section.

Fig.\,\ref{fig:exclusion_plot} shows the resulting constraints for the IMO scenario and the QD neutrino-mass spectrum in the left and right panels, respectively. We do not find appreciable sensitivity for NMO, since the scalar induced $\nu_e$ component is strongly suppressed by the small electron-flavour projection in this case. Hence, we do not show our results for NMO for all the three experimental setups.

The sensitivity exhibits a clear dependence on the scalar mass $\mp$. At small $\mp$, both the prompt and delayed neutrino spectra are concentrated at relatively low energies~(see Fig.~\ref{fig:DSNB_flux}, right panel), where the standard DSNB flux is largest. The scalar-induced excess is therefore more difficult to distinguish from the standard background. Increasing $\mp$ shifts both components toward higher energies, where the standard DSNB falls rapidly, resulting in stronger sensitivity. At sufficiently large values of $\mp$~($\mp \gtrsim 70\, {\rm MeV}$), the background from atmospheric muons and neutrinos starts to dominate over the scalar decay signal. As a result, improvement in sensitivity saturates. The hatched region is excluded by requiring that the energy released in the phase transition remain below the total neutrino energy radiated by the SN,  {\it i.e.} $E_{\rm cond} < E^{\rm tot}_\nu$, for the assumed core volume.

Remarkably, the present Super-K data already constrains the scalar parameter space corresponding to a FOPT for the IMO scenario as well as the QD cases. For $\mp \sim 100$ MeV, Super-K excludes $v_\phi > 56~{\rm MeV}~(63~{\rm MeV})$ at 95\% C.L for the IMO~(QD) case. To the best of our knowledge, this is the first study to investigate DSNB sensitivity to a first-order phase transition occurring in the scalar sector inside a CCSN.

As expected, the larger statistics expected from future detectors significantly improve this reach. With a 10-year exposure, DUNE can exclude $v_\phi \gtrsim 33~{\rm MeV}~(42~{\rm MeV})$ for $\mp \sim 100$~MeV at 95\% confidence level for the IMO~(QD) case. Hyper-K provides the strongest projected sensitivity, reaching $v_\phi \gtrsim 30~\mathrm{MeV}$~(36 MeV) for the IMO~(QD) scenario, for the benchmark exposure considered here.

Overall, the sensitivity is driven by the characteristic high-energy enhancement produced by the scalar-induced neutrino components. The complementarity between the prompt and delayed spectra, together with the different flavour sensitivity of DUNE and water-Cherenkov detectors, allows DSNB observations to probe a region of the parameter space in which the phase transition inside a CCSN can leave an observable imprint on the diffuse neutrino background. This demonstrates that CCSNe can provide an astrophysical laboratory for finite-temperature dynamics in weakly coupled scalar sectors.

\section{Concluding remarks}
\label{sec:conclusion}

The production of a new scalar through neutrino interactions can have important consequences for the neutrino emission from CCSN. Due to these interactions, the scalar can be trapped inside the SN core, and attain thermal equilibrium with the surrounding medium. For scalar masses up to $\mathcal{O}(100)$ MeV, the temperatures reached in the post-bounce SN core can be sufficiently high to induce a first-order phase transition in the scalar sector. In this work, we have explored the phenomenological consequences of such a phase transition for the DSNB.

The scalar sector gives rise to two distinct contributions to the neutrino flux from a SN. The first originates from the thermal scalar population that remains trapped in the SN and subsequently decouples from the medium, after which the scalars decay into neutrinos. We refer to this as the prompt neutrino component, since these decays occur on timescales comparable to the propagation of the scalar population out of the SN. The second contribution arises from the scalar field itself after the SN core cools below the critical temperature. The field then evolves toward the true vacuum and undergoes coherent oscillations about its vacuum expectation value. The subsequent decay of this condensate produces a delayed neutrino component. Since the condensate consists predominantly of non-relativistic scalar quanta, this component is characterised by a much narrower energy spectrum, with its characteristic energy set by the scalar mass.

We find that these scalar-induced contributions can significantly modify the standard DSNB spectrum. Their observable magnitude, however, depends strongly on the neutrino mass spectrum and flavour composition. For the NMO, the projection of the scalar-produced mass-eigenstates onto the electron flavour is strongly suppressed, substantially reducing the sensitivity of DSNB experiments. In the IMO, the scalar-induced contribution can be considerably larger, and the delayed component can exceed the standard DSNB flux over part of the $(\mp,v_\phi)$ parameter space. A similar enhancement occurs for a QD neutrino spectrum. Importantly, the delayed component is concentrated in an intermediate-energy region where the standard DSNB flux has already decreased, while the atmospheric-neutrino and atmospheric-muon backgrounds have not yet become dominant. This spectral separation provides a particularly favorable handle for identifying the scalar-induced contribution.

We have quantified the resulting sensitivity using the current DSNB results from Super-K and the projected sensitivities of DUNE and Hyper-K. For the NMO, the scalar-induced signal is too suppressed for the experimental configurations considered here to provide meaningful constraints. For IMO, however, the existing Super-K data already probe the scalar parameter space associated with the first-order phase transition, excluding values of $v_\phi\gtrsim 56$ MeV~(63 MeV) at 95\% confidence level for the IMO~(QD) case. Future experiments can substantially improve this reach. With a 10 year exposure, DUNE is expected to reach up to $v_\phi > 33$ MeV~(42 MeV), while the larger statistics of Hyper-K can extend the sensitivity to approximately  $v_\phi\gtrsim 30$ MeV~(36 MeV) for the IMO~(QD) scenario.

Our results demonstrate that the DSNB can serve as a novel probe of phase-transition dynamics occurring inside CCSN. In particular, the combination of a prompt thermal component and a delayed component provides a distinct spectral feature that can imprint itself upon the DSNB. This opens the possibility of using current and next-generation neutrino detectors to search for a FOPT happening at $\mathcal{O}(10-100)$ MeV, which could be a potential candidate for the stochastic gravitational wave (GW) signals in the nano-hertz (Hz) regime, reported by NANOGrav \cite{NANOGrav:2023gor,Manchester_2013,NANOGrav:2023hvm,Costa:2025csj,Goncalves:2025uwh,Balan:2025uke}. Our study shows that CCSNe can act as laboratories for finite-temperature scalar dynamics, with the DSNB providing an additional and potentially powerful avenue for probing first-order phase transitions in the dark or weakly coupled sector.

\subsubsection*{Acknowledgments}
We would like to thank Anna Suliga for useful discussions and Mary Hall Reno for the comments on the manuscript. SD is supported by the U.S. Department of Energy Grant DE-SC-0010113. MS and SN acknowledge support from the Early Career Research Grant by Anusandhan National Research Foundation (ANRF/ECRG/2024/000522/PMS).

\bibliographystyle{JHEP}
\bibliography{refs}

\end{document}